%% file: main.tex
\newif\ifdraft
\drafttrue

\documentclass[journal]{IEEEtran}

\usepackage{cite}
\usepackage{amsmath,amssymb,amsfonts}
\usepackage{algorithmic}
\usepackage{graphicx}
\usepackage{textcomp}
\usepackage{xcolor}
\usepackage[hidelinks]{hyperref}
\usepackage{listings}
\usepackage{textcomp}
\usepackage{siunitx}
\usepackage{multirow}
\usepackage{xspace}
\usepackage{balance}
\usepackage{hyperref}
\ifCLASSOPTIONcompsoc
    \usepackage[caption=false,font=normalsize,labelfont=sf,textfont=sf]{subfig}
\else
    \usepackage[caption=false,font=footnotesize]{subfig}
\fi



\usepackage{inconsolata}
\usepackage{comment}

\definecolor{backgroundColour}{HTML}{F8F8F8}
\definecolor{keywordclr}{HTML}{3969AC}
\definecolor{commentclr}{HTML}{A2A0AA}
\definecolor{stringsclr}{HTML}{11A579}
\definecolor{fnctionclr}{rgb}{0.467, 0, 0.533}
\definecolor{builtinclr}{rgb}{0.35, 0, 0.533}
\definecolor{symbolsclr}{rgb}{0.5, 0.25, 0.25}
\definecolor{numbersclr}{rgb}{0.8, 0.2, 0}
\definecolor{bckgrndclr}{rgb}{0.91, 0.95, 0.95}
\lstdefinestyle{PythonStyle}{
    language=Python,
    backgroundcolor=\color{backgroundColour},
    keywordstyle=\color{keywordclr}\bfseries,
    stringstyle=\color{stringsclr},
    commentstyle=\color{commentclr}\itshape,
    upquote=true,
    basicstyle=\ttfamily\linespread{0.9}\scriptsize,
    breakatwhitespace=false,
    breaklines=true,
    captionpos=b,
    keepspaces=true,
    numbers=left,
    numbersep=5pt,
    numberstyle=\color{commentclr}\ttfamily\tiny,
    showspaces=false,
    showstringspaces=false,
    showtabs=false,
    tabsize=2,
    xleftmargin=1.25em,
    frame=single,
    framexleftmargin=1.25em,
    morekeywords={assert,with,as,None}
}
\makeatletter
\apptocmd\@makecaption{\par}{}{%
  \errmessage{\noexpand\@makecaption could not be patched}%
}
\makeatother

\ifdraft
  \newcommand{\todo}[1]{{\textcolor{red}{ TODO: #1 }}}
  \newcommand{\ian}[1]{{\textcolor{green}{ Ian: #1 }}}
  \newcommand{\logan}[1]{{\textcolor{blue}{ Logan: #1 }}}
  \newcommand{\greg}[1]{{\textcolor{purple}{ Greg: #1 }}}
  \newcommand{\val}[1]{{\textcolor{orange}{ Valerie: #1 }}}
  \newcommand{\kyle}[1]{{\textcolor{teal}{ Kyle: #1 }}}
  
\else
  \newcommand{\todo}[1]{}
  \newcommand{\ian}[1]{}
  \newcommand{\logan}[1]{}
  \newcommand{\greg}[1]{}
  \newcommand{\val}[1]{}
  \newcommand{\kyle}[1]{}
  
\fi

\newcommand{\proxystore}{Proxy\-Store\xspace}

\begin{document}

\title{Object Proxy Patterns for Accelerating\\Distributed Applications}

\author{
J. Gregory Pauloski,
Valerie Hayot-Sasson,
Logan Ward,
Alexander Brace,\\
André Bauer,
Kyle Chard,
and Ian Foster,~\IEEEmembership{Fellow,~IEEE}
\thanks{
J. Gregory Pauloski, Valerie Hayot-Sasson, Alexander Brace, Kyle Chard, and Ian Foster are with the Department of Computer Science, University of Chicago, Chicago, IL, USA (e-mail: \href{mailto:jgpauloski@uchicago.edu}{jgpauloski@uchicago.edu}).
}
\thanks{J. Gregory Pauloski, Valerie Hayot-Sasson, Logan Ward, Alexander Brace, Kyle Chard, and Ian Foster are with the Data Science and Learning Division, Argonne National Laboratory, Lemont, IL, USA.
}
\thanks{
André Bauer is with the Department of Computer Science, Illinois Institute of Technology, Chicago, IL, USA.
}
}

\maketitle

\begin{abstract}
Workflow and serverless frameworks have empowered new approaches to distributed application design by abstracting compute resources.
However, their typically limited or one-size-fits-all support for advanced data flow patterns leaves optimization to the application programmer---optimization that becomes more difficult as data become larger.
The transparent object proxy, which provides wide-area references that can resolve to data regardless of location, has been demonstrated as an effective low-level building block in such situations.
Here we propose three high-level proxy-based programming patterns---distributed futures, streaming, and ownership---that make the power of the proxy pattern usable for more complex and dynamic distributed program structures.
We motivate these patterns via careful review of application requirements and describe implementations of each pattern.
We evaluate our implementations through a suite of benchmarks and by applying them in three meaningful scientific applications, in which we demonstrate substantial improvements in runtime, throughput, and memory usage.
\end{abstract}

\begin{IEEEkeywords}
Distributed Computing, Futures, Streaming, Memory Management, Open-source Software
\end{IEEEkeywords}

\section{Introduction}
\label{sec:introduction}

\IEEEPARstart{T}{ask-based} programming paradigms, such as function-as-a-service (FaaS) and workflows, have emerged as vital methods for achieving computational flexibility and scalability.
Applications are written as compositions of many distinct components, referred to as \textit{tasks}, and FaaS platforms and workflow systems, collectively referred to as \textit{execution engines}, abstract the complexities of executing tasks in parallel, whether across personal, cloud, edge, and/or high-performance computing (HPC) systems~\cite{rocklin2015dask,azurefunctions,googlecloudfunctions,amazonlambda,babuji19parsl,chard20funcx}.
Such execution engines have enabled a wide variety of innovative applications.

Yet as the scale and ambition of task-parallel applications grows, they increasingly encounter difficulties due to the use of shared storage for the exchange of intermediate data among tasks---an approach commonly employed both by workflow systems (e.g., Parsl~\cite{babuji19parsl}, Pegasus~\cite{deelman15pegasus}, Swift~\cite{wilde11swift}) and cloud-hosted FaaS systems (e.g., AWS Lambda~\cite{amazonlambda}, Azure Functions~\cite{azure2011cloudstorage}, Google Cloud Function~\cite{googlecloudfunctions}).
Such uses of shared storage can fail or become prohibitively expensive as the number of tasks, the geographic distribution of tasks, the quantities of data exchanged, and the required speed of data exchange grow.

Many researchers have investigated alternative mechanisms for distributed and wide-area data management that circumvent these limitations of shared storage.
For example, Linda's tuple space model provides unified access to a shared distributed memory space~\cite{linda}, DataSpaces provides a similar model for large-scale applications~\cite{docan2010dataspaces,dataspaces2017aktas}, and peer-to-peer systems like the InterPlanetary File System provide decentralized content-addressed file sharing~\cite{benet2014ipfs}.
Another approach to simplifying data sharing is the \textit{object proxy paradigm}, which provides transparent access and management for shared objects in distributed settings.
This mechanism, long used with Java's Remote Method Invocation (RMI)~\cite{biegel2002dynamic}, is also supported in Python via the ProxyStore system~\cite{pauloski2023proxystore}.
ProxyStore's transparent object proxies provide lightweight, wide-area references to objects in arbitrary data stores---references that can be communicated cheaply and resolved just-in-time via performant bulk transfer methods in a manner that is transparent to the consumer code.
Recent work has shown how by decoupling data flow complexities from control flow-optimized execution engines~\cite{ward2021colmena,harb2023uncovering,collier2023developing,dharuman2023protein,pauloski2024taps}, the object proxy paradigm can simplify implementations of dynamic application structures such as ML model training and inference.

\input{figures/patterns-overview}

Yet the object proxy paradigm remains a \textit{low-level} abstraction that can be hard to use in practice due to the complexities inherent in managing many references to remote objects.
Thus, we ask: Can we identify common \textit{high-level patterns} that build on the proxy model to accelerate and simplify development of advanced applications?
To this end, we review in this paper three computational science applications previously developed for workflow execution engines (\emph{1000 Genomes}, \emph{DeepDriveMD}, \emph{MOF Generation}), identify limitations in the data flow patterns supported by those execution engines, and
propose three new programming patterns that extend the proxy model to overcome these limitations (\autoref{fig:patterns-overview}):
\begin{itemize}
    \item A \emph{distributed futures} system for seamless injection of data flow dependencies into arbitrary compute tasks to overlap computation and communication;
    \item An \emph{object streaming} interface that decouples event notifications from bulk data transfer such that data producers can unilaterally determine optimal transfer methods; and
    \item An \emph{ownership} model that provides client-side mechanisms for managing object lifetimes and preventing data races in distributed task-based workflows.
\end{itemize}

Each pattern simplifies building sophisticated task-based applications that are to execute across distributed or remote compute resources (e.g., using FaaS or workflow systems).
For each, we discuss its requirements and the protocols used to support it.
Our reference implementations extend \proxystore{}~\cite{pauloski2023proxystore}, our prior work, to leverage the existing low-level proxy model within Python, a popular and pervasive language for task-based distributed applications.
The implementations are available within \proxystore{} v0.6.5 and later, available on GitHub~\cite{proxystore-github} and PyPI~\cite{proxystore-pypi}.
We evaluate our reference implementation for each pattern using (1) synthetic benchmarks across various FaaS and workflow systems and (2)~our motivating applications, for which we reduce workflow makespan by 36\% in 1000 Genomes, improve inference latency by 32\% in DeepDriveMD, and optimize proxy lifetimes in MOF Generation.

The rest of this paper is as follows:
\autoref{sec:motivation} introduces our motivating scientific applications;
\autoref{sec:background} gives background on our prior work, ProxyStore;
\autoref{sec:patterns} outlines the design and implementation of each pattern;
\autoref{sec:evaluation} demonstrates synthetic evaluations;
\autoref{sec:applications} presents our experiences applying these patterns to our motivating applications;
\autoref{sec:related} provides context about related work; and
\autoref{sec:conclusion} summarizes our contributions and future directions.

\section{Motivating Applications}
\label{sec:motivation}

\textbf{1000 Genomes:}
This bioinformatics pipeline~\cite{1000genomesworkflow} identifies mutational overlaps within the \num{2504} human genomes sequenced by the 1000 Genomes Project~\cite{1000genomesproject}.
It comprises five stages:
(1) fetch files, each containing all Single Nucleotide Polymorphisms (SNPs) in a chromosome, chunk, and process them in parallel to extract SNP variants by individual;
(2) merge individuals' results of the prior stage;
(3) score and select SNP variants based on their phenotypic effect;
(4) compute overlap of selected SNP variants among pairs of individuals and by chromosome; and
(5) compute frequency of overlapping variants.
Executing scientific workflows in a FaaS setting may be preferred when access to specialized hardware, such as AI or quantum accelerators, or the ability to rapidly scale up or down is required, but workflow execution on a FaaS system poses challenges because FaaS systems rely on control flow to determine when to submit tasks.
From the application perspective, however, the availability of data---the data flow---is the condition upon which tasks can be submitted.
We use the 1000 Genomes workflow as an example of the challenges that arise when executing data flow oriented applications on control flow-optimized systems.

\textbf{DeepDriveMD:}
Molecular dynamics (MD) simulation acts as a computational microscope~\cite{dror2012biomolecular} to enable the study of complex biomolecular systems.
However, many important phenomena are difficult to sample using conventional MD, even with powerful supercomputers~\cite{hospital2015molecular}.
DeepDriveMD~\cite{lee2019deepdrivemd, brace2022deepdrivemd} implements an emerging HPC paradigm in which machine learning (ML) methods are used to track a simulated state space and guide simulations toward a sampling objective.
The DeepDriveMD client submits discrete training, inference, and simulation tasks and receives their results.
This pattern causes two challenges.
First, all data must flow through the client which limits performance at scale (e.g., data volume or task frequency), so a mechanism is needed to alleviate data flow burdens from the client when possible.
Second, repeated tasks perform redundant work.
For example, each inference tasks loads the latest ML model from disk, infers using the input batch, and compiles the results which will later become the input to a simulation task.
This is inefficient because the same model is loaded multiple times across tasks, tasks may execute on different workers negating cache benefits, and every task incurs non-trivial overheads for scheduling and execution.

\textbf{Metal-Organic Framework (MOF) Generation:}
This workflow~\cite{mofgithub} uses molecular diffusion models~\cite{park2024mof} to generate organic ligands, assemble MOF candidates, and employ physics models to identify candidates best suited for storing CO$_2$.
The workflow uses a central process, referred to as a thinker~\cite{ward2021colmena}, to determine which tasks to execute, and with what parameters.
A core computational challenge is ensuring that the thinker has timely data, such as the latest diffusion model results, when deciding the next task.
Object proxies have been used to improve thinker response time in similar applications~\cite{ward2021colmena,ward2023colmena}, but knowing the lifetime of proxied data is challenging in sophisticated workflows where the types of tasks to be executed are not know ahead of time.

\section{ProxyStore}
\label{sec:background}

In software design, a \emph{proxy} is an object that functions as an interface to another object~\cite{gamma1994design}.
A simple proxy will forward operations on itself to the real or \emph{target} object, but often a proxy is used to provide extra functionality such as caching or access control, in addition to forwarding operations~\cite{pauloski2023proxystore}.
For example, distributed applications can use a proxy to invoke methods on a remote object, and data-intensive applications can use a virtual or \emph{lazy} proxy which will perform just-in-time resolution of large objects (i.e., load the object from a remote location into local memory when first needed).

Lazy transparent object proxies can be used to communicate objects efficiently in distributed applications~\cite{pauloski2023proxystore}.
Here, a proxy refers to a target object stored in an arbitrary mediated communication medium (e.g., an object store, database, file system).
The proxy forwards all operations on itself to the target, but importantly is totally transparent in that the proxy is an instance of the same type as the target.
In Python, this means that \texttt{isinstance(p, type(t))} is true for a proxy \emph{p} and its target \emph{t}.
The proxy is lazy in that it performs just-in-time resolution of the target.
The target is not copied from the mediated storage into local memory until an operation is invoked on the proxy.
This proxy paradigm has both pass-by-reference and pass-by-value semantics; unused copies of the target object are not made when the proxy is passed between processes but the actual consumer of the proxy is given a copy.

The benefits of moving data via proxies are numerous: pass-by-reference reduces transfer overheads, no external information is required to resolve a proxy, shims or wrapper functions are eliminated, just-in-time resolution amortizes communication costs and avoids costs associated with unused objects, and proxies enable automatic access control.
As such, this paradigm has been used to build a diverse suite of robust and scalable scientific applications~\cite{ward2021colmena,zvyagin2023genslms,ward2023colmena,collier2023developing,harb2023uncovering,kamatar2023lazy,pauloski2023proxystore,dharuman2023protein}.

\input{figures/patterns-proxystore-stack}

\proxystore{}~\cite{pauloski2023proxystore} implements this proxy paradigm which we use as the basis for our patterns' reference implementations.
\proxystore{} defines the \emph{factory}, \emph{connector}, and \emph{store} constructs.
The \emph{factory} is a callable object that returns the target object when invoked.
\proxystore{} creates a unique factory for each target object containing the metadata and logic necessary to retrieve the target from a remote location.
This factory is used to initialize a proxy, and a proxy is \emph{resolved} once it has invoked its factory to retrieve and cache the target locally.

The \emph{connector} is a protocol that defines the low-level interface to a mediated communication channel.
A mediated channel is one where the communication between a producer and consumer is indirect, such as via a storage system~\cite{copik2022fmi}.
This indirection is important because the process that creates a proxy and the process that resolves a proxy may not be active at the same time, in which case they could not communicate via direct mechanisms.
\proxystore{} provides many connectors, including interfaces to external mediated channels such as shared file systems, object stores (Redis~\cite{redis} and KeyDB~\cite{keydb}), and peer-to-peer transfer systems (Globus Transfer~\cite{foster2011globus,chard2014globus} and \proxystore{} Endpoints~\cite{pauloski2023proxystore}) and bespoke mediated channels that can leverage high-performance networks through the UCX~\cite{UCX-Py} and Margo~\cite{py-mochi-margo} libraries.

The high-level \emph{store} interface, initialized with a connector, is used to create proxies of objects.
A proxy \emph{p} can be created from a target \emph{t} by calling \texttt{Store.proxy(t)}.
This method (1)~serializes \emph{t} using the default ProxyStore or user-provided serializer~\cite{pauloski2024accelerating}, (2) puts the serialized \emph{t} in the mediated channel via the connector, (3) creates a factory with the appropriate metadata about \emph{t} and the store/connector used, (4) initializes a proxy with the factory, and (5) returns the proxy.
This process incurs some overhead but is trivial for larger objects.
Prior work~\cite{pauloski2023proxystore,ward2021colmena} found the performance benefits of proxies to outweigh proxy creation and resolution overhead for objects larger than $\sim$10~kB; the exact threshold depends on many factors (e.g., connector choice, execution engine).

\section{Proxy Patterns}
\label{sec:patterns}

We describe the design of each of the three advanced programming patterns that build on the aforementioned distributed object proxy base.
We discuss the details of our reference implementations that extend \proxystore{}, and \autoref{fig:patterns-proxystore-stack} describes how these patterns fit into the existing \proxystore{} stack.
These patterns are not mutually exclusive, but we discuss each in isolation for clarity.

\subsection{Distributed Futures}
\label{sec:patterns:futures}

\input{figures/task-pipelining}
\input{listings/futures-example}

A future represents a value that will eventually be available; the holder of a future can block on it until the value is resolved.
Futures simplify writing non-blocking compute (e.g., remote procedure calls, database queries, or FaaS invocations) and I/O (e.g., network requests or file system reads) operations.
Execution engines use futures to represent eventual task results, and this is valuable for representing long running remote execution or assembling applications with asynchronous callbacks.
However, the distributed futures provided by execution engines have three key limitations:
(1) these futures perform control and data synchronization so data flow cannot be optimized independent of control flow, such as to pipeline task execution as in \autoref{fig:task-pipelining};
(2) the transfer mechanisms used by the future cannot be optimized based on the type or location of data; and
(3) futures produced by execution engines are only usable within the context of that execution engine (e.g., a future from one engine cannot be sent as input to another).

We design a distributed futures system called \emph{ProxyFutures} that (1) supports explicit and implicit usage, arbitrary execution engines, arbitrary distributed memory backends, and seamless injection of data flow dependencies, and (2) addresses a limitation of \proxystore{} that a proxy cannot be created before its target object exists.
In ProxyFutures, a future $f$ is created for an eventual value $x$, and $f$ can be used to create any number of proxies $p_f$.

Consider an application with a main process $M$, a data producing process $P$, and a data consuming process $C$.
$M$ dispatches two tasks: $T_P$ to $P$ and $T_C$ to $C$.
$T_P$ is to produce a value $x$ to be consumed by $T_C$; thus,
$T_C$ has a data dependency on $T_P$.
$M$ can create a future $f$ and associated proxy $p_f$, and pass $f$ and $p_f$ to $T_P$ and $T_C$, respectively.
When $T_C$ first resolves $p_f$, it blocks until $T_P$ has set the result of $f$.
Importantly, $T_C$ can be started before $T_P$ has finished or even started.
$M$, when creating $f$, can choose the communication method to be used based on where $P$ and $C$ are located and what communication methods are available between them;
thus, the detailed communication semantics are abstracted from $T_P$ and $T_C$.
The implicit nature of $p_f$ also means that the code for $T_C$ can be invoked either on a value directly or on a proxy of the value.
This equivalence simplifies code and testing and means that $M$ can inject data flow dependencies via a future into arbitrary third-party functions that expect to receive data directly.

We implement this behavior by extending \proxystore{}'s \texttt{Store} interface to expose a \texttt{future()} method that returns a \texttt{Future} object.
The \texttt{Future} class exposes two main methods: \texttt{set\_result(obj: T)}, which sets the result of the future to an object of type \texttt{T} and \texttt{proxy()}, which returns a \texttt{Proxy[T]}.
When a proxy created via \texttt{Future.proxy()} is resolved, the proxy blocks until the target value has been set via a call to \texttt{Future.set\_result()}, as shown in \autoref{lst:futures-example}.
Use of ProxyFutures does not affect when a successor task starts; scheduling is still managed by the execution engine and/or user application.
ProxyFutures are best integrated at the application level so that developers can optimize task execution per their application requirements and to express more complex data dependencies than typically supported by execution engines.

Internally, communication between a \texttt{Future} and any child proxy(s) is handled via the \texttt{Store} used to create the \texttt{Future}.
Thus, a future and associated proxies can be serialized and sent to arbitrary processes on arbitrary machines.
In contrast, many standard-library future implementations use non-serializable async, thread, and inter-process synchronization mechanisms (e.g., \texttt{std::future} in C++~\cite{cpp-std-futures}, concurrent and async futures in Python~\cite{python-futures}), while RPC-based futures are only resolvable within the RPC framework (Dask futures~\cite{rocklin2015dask} or Ray \texttt{ObjectRefs}~\cite{wang2021ownership}).
The self-contained properties of the proxy mean that all logic for communication and resolution are embedded within the future and proxy; the future creator chooses communication methods on behalf of the process(es) which might set or consume the result of the future.

\subsection{Object Streaming}
\label{sec:patterns:streaming}

\input{figures/proxystore-streaming}
\input{listings/streaming-globus-compute}

High-performance stream processing applications dispatch remote compute tasks on objects consumed from a stream, but task dispatch can quickly become a bottleneck with high throughput streams~\cite{lee2019deepdrivemd, brace2022deepdrivemd, zvyagin2023genslms}.
Consider the application in~\autoref{fig:proxystore-streaming}, where process $A$ is a data generator that streams chunks of data (i.e., arbitrary Python objects) to a dispatcher process $B$, which for each data chunk dispatches a compute task on a remote process $C$.
Note that while the dispatcher consumes from the stream, it does not need the actual chunk of data; rather, it only needs to know that a chunk is ready (and potentially have access to user-provided metadata) in order to dispatch the task that will actually consume the chunk.
We design a system called \emph{ProxyStream} to enable scalable applications of this pattern.
At its core, ProxyStream uses a stream of proxies, rather than data chunks.
Bulk data are only transmitted between the data generator and the process/node computing on the proxy of the chunk, bypassing the intermediate dispatching process.
ProxyStream optimizes for both metadata and bulk data transfer, has broad execution engine compatibility, provides a self-describing data format, and supports various communication modules to take advantage of high-performance networking stacks.

ProxyStream provides two high-level constructs, the \texttt{Stream\-Producer} and \texttt{Stream\-Consumer}, that combine a message stream broker for low-latency event metadata propagation and a mediated communication channel for efficient bulk data transfer.
A \texttt{StreamProducer} is initialized with a \texttt{Publisher} and a \proxystore{} \texttt{Store}.
The \texttt{Publisher} defines a protocol for sending event messages to a stream.
We provide shims to many popular event streaming systems (Kafka~\cite{kafka}, Redis Pub/Sub and Queues~\cite{redis}, ZeroMQ~\cite{hintjens2013zeromq}) which implement the \texttt{Publisher} protocol.
When a new object and optional metadata are sent to the \texttt{StreamProducer}, (1) the object is put in the store, (2) a new event containing the user provided metadata and information about where the object is stored is created, and (3) the event is published via the \texttt{Publisher}.

A \texttt{StreamConsumer} is initialized with a \texttt{Subscriber}, which, like the \texttt{Publisher}, defines a protocol for receiving event messages from a stream (\autoref{lst:streaming-gc}).
The \texttt{StreamConsumer} is an iterable object, yielding proxies of objects in the stream until the stream is closed.
Calling \texttt{next()} on the \texttt{StreamConsumer} waits for a new event metadata message via the \texttt{Subscriber}, creates a proxy of the object using the event metadata, and returns the proxy to the calling code.
This process is efficient because the bulk object data has not been read at this point; rather, this will be delayed until the resolution of the proxy.

This model has many benefits: (1) communication mechanisms are abstracted from the stream consumer, (2) stream objects are resolved only when actually needed (wherever the proxy is resolved), (3) event message and bulk data transfer are decoupled, allowing the application to better optimize both forms of communication for the given application deployment environment and object characteristics, and (4) it provides a mechanism for implementing stateful actors in a workflow.

The ProxyStream interfaces support any combination of single/multi producer/consumer that is
supported by the associated \texttt{Publisher} and \texttt{Subscriber} implementations.
The \texttt{StreamProducer} supports mapping different stream topics to \texttt{Store} instances, enabling further optimization of communication mechanisms; batching; and plugins for filtering, sampling, and aggregation.
The \texttt{StreamConsumer} support plugins for filtering and sampling.
ProxyStream is fault-tolerant provided that the broker and communication channel are fault-tolerant.

ProxyStream can be integrated at the application or framework level.
\autoref{lst:streaming-gc} depicts use of ProxyStream within a Globus Compute application; we integrate ProxyStream within the DeepDriveMD framework for the evaluation in \autoref{sec:applications}.

\subsection{Ownership}
\label{sec:patterns:ownership}

A limitation of the proxy model is the need to manage explicitly the lifetime of the associated target object.
When a proxy is shared with more than one process, it is challenging to know when it is safe to free the target object.
A \proxystore{} proxy acts like a C/C++ pointer or raw pointer in Rust; thus, one process could prematurely free the target object, causing what is equivalent to a null pointer exception in the other process(es); delay freeing the object causing increased memory usage; or forget to free the object causing a memory leak.
\proxystore{} provides some guidance on using proxies safely, but ultimately it is up to the programmer to use proxies safely---a situation similar to C pointers.

To address this difficulty, we extend the proxy model with two features not provided by \proxystore{}: automatic deletion of objects that have gone out of scope and safe support for mutating objects.
Inspired by Rust's borrowing and ownership semantics, our design works in distributed contexts; provides different proxy types that can represent the owned, reference, and mutable reference types; enforces ownership and borrowing rules at runtime based on a proxy’s type; and performs automatic dereferencing, coercion, and deletion.

Rust defines three \textit{ownership rules}: (1) each value has an owner, (2) there can only be one owner at a time, and (3) a value is deleted when its owner goes out of scope~\cite{rustlang}.
A reference allows a value to be borrowed without relinquishing ownership.
The \textit{reference rules} are (1) at any given time, a value can have either one mutable reference or any number of immutable references and (2) references must always be valid.
The Rust compiler enforces these rules, and the language provides data structures for runtime enforcement for more complex scenarios that the compiler cannot reason about.

Applying these rules in a distributed application, such as a computational workflow, can make memory management significantly easier without the need to perform global reference counting.
Computations represented as directed acyclic graphs (DAGs) are particularly well suited to this model.
As objects move from a parent DAG node to a child node, ownership can either be transferred to the child or the child can be given a borrowed reference.
Thus, a node has full information about what operations are safe on objects that it receives.
Ownership transfer means that the recipient node has full control over that object;
an immutable reference means that the node can only read the object.
A mutable reference means that the node has sole access to modify the object, but the node cannot create and share additional references: i.e.,
it is not allowed to pass a reference to its own child node.

One challenge of this model is knowing when a reference to an object goes out of scope, because this requires communication between the process that owns the object and the process that has a reference.
However, in a task-based workflow, it is easy to reason that a reference passed to a task goes out of scope when the task completes (assuming that the task is well-behaved; an improperly behaved task would be one that, for example, creates and stores a memory-to-memory copy of the reference) and workflow systems already propagate information about task completion.

A second challenge is representing the ownership or borrowing of an object.
The Rust compiler and dot operator abstracts much of the nuance of dealing either with objects directly or with their references~\cite{rustdotoperator}.
In Python, for example, an object \texttt{T} could be wrapped in a \texttt{Owned[T]}, \texttt{Ref[T]}, and \texttt{RefMut[T]}, in a similar manner to some Rust constructs.
However, use of these constructs would be cumbersome, as all referencing, dereferencing, or coercion would have to be done manually.

The transparent object proxy is well-suited to solve these object scope and reference representation problems.
An object that is proxied by a process becomes a shared object that is stored on some global object store accessible by all processes in the distributed environment (\autoref{lst:ownership-stubs}).
The target object is serialized, put in the global store, and an \texttt{OwnedProxy} is returned.
The \texttt{OwnedProxy} contains a reference to the global object and, if the proxy has been resolved, a local copy of the object upon which the proxy forwards operations to.

An \texttt{OwnedProxy} enforces the following rules [c.f.\ Rust’s ownership rules]: (1) each object in the global store has an associated \texttt{OwnedProxy}, (2) there can only be one \texttt{OwnedProxy} for any object in the global store, and (3) when \texttt{OwnedProxy} goes out of scope, the object is removed from the global store.

\input{listings/ownership-stubs}

When invoking a task on an \texttt{OwnedProxy} (i.e., calling a local or remote function), the caller can do one of four things:
\begin{itemize}
    \item Yield ownership by passing the \texttt{OwnedProxy} to the task.

    \item Clone \texttt{OwnedProxy} and pass the cloned \texttt{OwnedProxy} to the task.
    Cloning an \texttt{OwnedProxy} will create a new copy of the object in the global store that will be owned by the callee task while the caller still owns the original object.

    \item Make a \texttt{RefProxy} and pass the \texttt{RefProxy} to the task.
    The caller still retains ownership, and the task can only read the object via the \texttt{RefProxy}.
    The callee task can only mutate its local copy, not the global copy.
    The caller's \texttt{OwnedProxy}, used to create the \texttt{RefProxy}, keeps track of the references that it has created.
    Any number of tasks can be invoked on a \texttt{RefProxy} at a time.

    \item Make a \texttt{RefMutProxy} and pass the \texttt{RefMutProxy} to the task.
    The caller still retains ownership (essentially the privilege to delete), but the callee task now has sole access to modify the object in the global store.
    The caller’s \texttt{OwnedProxy} marks that it has created a \texttt{RefMutProxy} and thus cannot mutate itself until the callee task that has the \texttt{RefMutProxy} completes.
    Only one task can be invoked on a \texttt{RefMutProxy} at a time and a \texttt{RefMutProxy} and \texttt{RefProxy} cannot exist at the same time.
\end{itemize}

The lifetimes of a \texttt{RefProxy} and \texttt{RefMutProxy} are strongly coupled to those of the tasks they are passed to.
Any violation of these rules, such as an \texttt{OwnedProxy} that goes out of scope or is deleted while a \texttt{RefProxy} or \texttt{RefMutProxy} exists, will raise a runtime error.
It is also possible to extend a static code analysis tool to verify correctness prior to execution.

Execution engines typically use futures to encapsulate the asynchronous execution of a task.
Thus, we use callbacks on the task result futures to indicate that the references associated with a task have gone out of scope.
The primary limitation of this approach is that each execution engine has a different syntax for submitting a task and getting back a future.
Rather than modify each engine, we provide a set of shims that appropriately parse task inputs and construct a callback on the task’s future that will propagate the necessary information about references going out of scope.
The \texttt{StoreExecutor}, an interface provided by ProxyStore, wraps an execution engine client (e.g., a Globus Compute, Dask, or Parsl client) and automatically proxies task parameters and results based on user-defined policies and manages references associated with tasks~\cite{pauloski2024accelerating}.
The \texttt{StoreExecutor} is easy to use, but applications requiring more fine-grain control can use the API in \autoref{lst:ownership-stubs}.

The ownership model is not fault-tolerant when the client crashes in a manner which prevents garbage collection, but the model is compatible with fault-tolerant execution engines such as those that automatically rerun tasks on failure.
Since only a single \texttt{RefMutProxy} can exist, the ownership model is not optimal for applications with many concurrent writers to the same object; a database, for example, may be more suitable.

So far, we have constricted ourselves to tasks (i.e., function invocations) as the only region of code over which we can define a lifetime; thus, all references to an object are equal to the lifetime of the single task invoked on that reference.
Yet a workflow application may employ more complex lifetimes.
For example, a lifetime could be assigned to a set of tasks that are a subgraph of the global DAG, and a programmer might want to define references to global objects that are associated with this custom lifetime.
Using proxy references is a valid solution but would require additional code to manage and map references to the scopes contextual to the application.

\input{listings/lifetimes}

We provide the \texttt{Lifetime} construct, an alternative to proxy references, for managing object lifetimes in more complex scenarios.
A lifetime, attached to one or more proxies upon proxy creation, will clean up associated objects once the lifetime has ended.
We provide three \texttt{Lifetime} types and the API can be extended to implement new types.
The context-manager lifetime enables mapping proxy lifetimes to discrete segments of code, the time-leased lifetime will clean up associated objects once the lease has expired and not been extended, and the static lifetime persists objects for the remainder of the program.
\autoref{lst:lifetimes} provides a time-leased lifetime example.

\section{Evaluation}
\label{sec:evaluation}

\newcommand{\evalsystem}{Polaris\xspace}

We conducted experiments on Polaris at the Argonne Leadership Computing Facility.
Polaris has 560 nodes interconnected by an HPE Slingshot 11 network and a 100 PB Lustre file system.
Each node contains one AMD EPYC Milan processor with 32 physical cores, 512~GB of DDR4 memory, and four 40~GB NVIDIA A100 GPUs.

\subsection{Task Pipelining with ProxyFutures}

\input{figures/task-pipelining-results}

We first evaluate the effectiveness of ProxyFutures for reducing workflow makespan via pipelining.
We define a synthetic benchmark that submits $n$ tasks in sequence, each sleeping for $s$ seconds and then producing $d$ bytes to be consumed by the next task.
As in \autoref{fig:task-pipelining}, a fraction $f$ of each task is treated as startup overhead (e.g., library loading, model initialization, state synchronization).
Thus, each task sleeps for $f\times s$ seconds, resolves its input data, and then sleeps for the remaining $(1-f)\times s$ seconds to simulate computation.
We compare three deployments: sequential without proxies (\textit{No Proxy}), sequential with proxies (\textit{Proxy}), and pipelined with ProxyFutures (\textit{ProxyFuture}).
In the first two, task $t_i$ is submitted once the result of task $t_{i-1}$ is available, with
in \textit{No Proxy}, the workflow engine handling data transfer, and in \textit{Proxy}, data transfer being offloaded from the workflow engine.
In \textit{ProxyFuture}, tasks $t_{i-1}$ and $t_i$ share a proxy and future pair and $t_i$ is submitted before $t_{i-1}$ is complete.

\textbf{Setup:}
We run a Dask cluster on a single \evalsystem compute node.
In the \textit{Proxy} and \textit{ProxyFuture} deployments, a Redis server running on the compute node is used as the mediated communication channel for the proxies.
We run $n$ = 8 tasks with intermediate data of $d$ = 10~MB and task time of $s$ = 1~s;
the short task time is to focus on the time spent producing and waiting on data.
We vary overhead fraction $f$ from 0 to 0.9.

\textbf{Results:}
We plot in \autoref{fig:task-pipelining-timeline} the start and end times of each stage in each task's lifecycle for each deployment, for $f=0.2$, and for \textit{ProxyFuture}, also for $f=0.5$.
Each task incurs fixed \textit{overhead} and \textit{compute} costs, of $f$ and (1~$-f$)~s, respectively.
Other costs include: \textit{submit}, the time to submit and begin execution; \textit{generate}, the time to produce output data; and \textit{receive}, the time to receive the result by the client.
\textit{Proxy} and \textit{ProxyFuture} also incur \textit{resolve} costs associated with the use of proxies.
\autoref{fig:task-pipelining-trend} shows the implications of these differences by presenting average makespan as a function of task overhead fraction for the three deployments.
The use of proxies in \textit{Proxy} improves task submission time relative to \textit{No Proxy}, reducing makespan by 12\%.
The pipeline overlapping in \textit{ProxyFuture} enables close to the theoretical limit (dashed line) as determined by inter-task data dependencies.
For example, the ideal makespan reduction of a pipeline execution is 20\% when $f$ = 0.2; we observe 19.6\% in \emph{ProxyFuture}.
The increased divergence from the ideal reduction at larger overhead fractions occurs because task submission and data transfer costs become more significant as overlapping increases.
Thus, a subsequent task begins waiting on its future slightly before the prior task has set the result of the future.

\textbf{Outcomes:}
DAG-based workflow execution models limit optimization of task execution because a child task cannot start until its parents have finished, even if the programmer knows it may be beneficial to start it sooner.
For example, module loading can account for a significant portion of overall task runtime.
Loading TensorFlow on NERSC's Perlmutter takes 5~s in the best case but nearly a minute when many workers read files concurrently~\cite{kamatar2023lazy}.
This is particularly noticeable with smaller models where inference time can be measured in fractions of a second.
On \evalsystem, the machine used here, we found that five common libraries (NumPy, Scikit-learn, SciPy, PyTorch, TensorFlow)~\cite{bauer2024globuscomputedataset} require from 100~ms to 2~s to import even under ideal conditions with a single worker.
Tasks must also often perform other work, such as file loading, initializing model weights, or state synchronization, before needing their input data.
The ProxyFutures model provides for seamless encoding of data dependencies and optimistic task pipelining when tasks have nontrivial initial overheads.
While we used Dask and Redis in this experiment, our approach will work with \emph{any} task-based execution engine and mediated communication channel.
This engine-agnostic approach will enable programmers to coordinate tasks across multiple execution engines concurrently.

\subsection{Scalable Stream Processing}

\input{figures/streaming-results}

Here we evaluate scalable stream processing with Proxy\-Stream.
As in \autoref{fig:proxystore-streaming}, there is one data producer publishing data of size $d$ to the stream with a rate $r$ (items per second).
A dispatch node consumes data from the stream and dispatches a compute task for each data item on to a cluster of $n$ workers.
Each compute task is simulated by a task which sleeps for $s$~seconds.
The dispatcher executes on a login node, and given $n$ workers, one worker is allocated as the producer while the remaining $n-1$ workers are used to execute compute tasks.

\textbf{Setup:}
We compare three streaming configurations.
In \emph{Redis Pub/Sub}, data are published directly to a Redis pub/sub topic that is consumed by the dispatcher before being sent to a worker to be computed on.
In \emph{ADIOS2}, data are written step-by-step to an ADIOS2 stream~\cite{godoy2020adios2}.
The dispatcher iterates on steps and launches worker tasks which will read the data from the ADIOS2 stream at a specified step.
In \emph{ProxyStream}, data are published to a \texttt{StreamProducer} which decouples metadata from bulk data, sending metadata to a Redis Pub/Sub topic and storing bulk data in a Redis Key/Value store.
The dispatcher consumes proxies of stream data via the \texttt{StreamConsumer} and sends proxies to workers to be computed on.
ADIOS2 and ProxyStream avoid data transfers through the dispatcher.

We use Parsl's \texttt{HighThroughputExecutor}, which can scale to thousands of tasks per second, to manage task execution.
We set the producer's data publishing rate $r=$ ($n-1)/s$ items per second, where $s=$ 1~s for all tasks.
Assuming no overheads in the system, this rate would keep each of the $n-1$ compute workers constantly fed with new data.
A range of data sizes $d$ and workers $n$ are evaluated to understand stream scaling throughput limitations.
We assign one worker per core so there are at most 32 workers per node.
We run each configuration for between five and thirty minutes, depending on the scale, which is long enough for the processing throughput (i.e., tasks completed per second) to stabilize.

\textbf{Results:}
\autoref{fig:streaming-results} shows the average compute tasks completed per second.
At the smallest data size, $d=100$~kB, performance is comparable between the three methods because data are not large enough to stress the system.
For larger worker counts $n$ and data sizes $d$, the default Redis Pub/Sub deployment slows because the dispatcher becomes a bottleneck, processing stream data at $\sim$100~MB/s.
This rate is slower than the network connection between the Redis server and dispatcher because the dispatcher must, for each stream item, receive and deserialize the item from Redis; compose the task payload, serializing the item again; and communicate the task payload to a worker.
Thus the dispatcher cannot process the incoming stream data fast enough to keep workers fed with new tasks when the number of workers or data size is sufficiently high.

ADIOS2 performs better than Redis Pub/Sub because we configured workers to read items from the stream directly based on a step index provided by the dispatcher, improving the latency between the dispatcher receiving stream data and launching a new task.
However, ADIOS2 requires changes to the worker task code not needed by the other two methods.

ProxyStream also alleviates data transfer and serialization burdens from the dispatcher enabling performance on par with or better than ADIOS2 but does so transparently without needing changes to the worker task code.
The peak processing throughput of ProxyStream is $1.7\times$ and $2.0\times$ faster than ADIOS2 for 1~MB and 10~MB item sizes, respectively.
Compared to the Redis Pub/Sub baseline, ProxyStream is $4.6\times$ and $6.2\times$ faster for 1~MB and 10~MB item sizes, respectively.
At $d=100$~MB, the largest data size evaluated, and $n=256$, ProxyStream is $7.3\times$ faster than Redis Pub/Sub.
ProxyStream and ADIOS2 perform similarly at this scale because other aspects of the experimental configuration become bottlenecks.
Namely, task execution overheads and storing the data produced by the generator limit peak throughput.
A faster data storage system or multiple data generators would be needed to achieve scaling beyond this point, and ProxyStream does support modular data storage and multi-producer configurations.

\textbf{Outcomes:}
Streaming proxies, rather than data directly, ensures that objects in the stream are only resolved once needed, thus avoiding overheads due to objects passing via intermediate processes.
The \texttt{StreamProducer} and \texttt{StreamConsumer} interfaces provide a mechanism for composing arbitrary message brokers and mediated communication methods, permitting developers to optimize application deployments without altering task code.
The resulting distributed applications are more portable and generalizable to new hardware systems.

\subsection{Memory Management}
\label{sec:evaluation:memory}

\input{figures/ownership-results}

We evaluate the automatic memory management of the proxy ownership model by comparing system memory usage over a simulated workflow to \proxystore{}'s default memory management and a manual memory management approach which relies on the a priori knowledge of the programmer to free shared objects.
We also compare to a baseline without any proxies where data are sent directly along with task requests.

\textbf{Setup:}
We execute a simulated workflow that imitates a series of map-reduces across a local Dask cluster on a single compute node of \evalsystem.
We run the workflow using each of the proxy memory management models, default, manual, and ownership, and a baseline without proxies using Dask for all data management.
We record average memory usage across three workflow executions for each configuration.
Eight consecutive map-reduces are performed where each of 32 mappers receives 100~MB and produces 10~MB.
We choose 100~MB because the value is large enough to be observable in the memory trace (i.e., larger than the baseline memory usage fluctuations) but is also below the Redis default maximum value size of 512~MB.
A single reducer consumes data produced by all mappers.
In addition to consuming and producing data, each tasks sleeps for 5~s.

\textbf{Results:}
\autoref{fig:ownership-results} presents the system memory usage traces for each memory management model.
The limited default memory management of \proxystore{} results in memory usage slowly increasing throughout execution as shared objects are created but never freed.
The automated management of our ownership model performs identically to manual management and appropriately evicts objects as references go out of scope.

The ``no proxy" baseline passes data directly to Dask and utilizes Dask's built-in distributed memory management.
We observe that Dask appropriately frees all task data; however, the overall runtime is three times slower.
The severe slow down is because Dask's graph serialization performs poorly with large ($>$1~MB in our experience) arbitrary Python objects.
Dask is optimized for transferring arrays and dataframes, and we found Dask's performance to be similar to the proxy cases when data were formatted as NumPy arrays.

\textbf{Outcomes:}
Our ownership model presents a marked improvement in using proxies in distributed workflows.
Enforcing ownership rules at runtime makes it easy to reason about what operations on shared objects are safe and prevents programming mistakes which may lead to memory leaks.
Our reference implementation is designed to be agnostic to the underlying task execution engine, but we believe that incorporating this model directly into execution engines can enable more powerful features.

\section{Applications}
\label{sec:applications}

\input{figures/1000-genomes-results}

\textbf{1000 Genomes:}
We use the 1000 Genomes workflow to investigate ProxyFutures as a mechanism for reducing task overheads and extending data flow dependencies to FaaS systems.
Tasks in the original 1000 Genomes workflow were implemented as Bash scripts.
We use the Python implementation of 1000 Genomes, where tasks are implemented as functions, to execute the workflow using a FaaS execution engine such as Globus Compute (which we use in the experiments reported here, due to its integration with HPC systems).

We evaluate the makespan of the resulting workflow, using 5\% of the 1000 Genomes dataset, on a single \texttt{compute-zen-3} node, with two 64-core CPUs and 256 GB memory, on Chameleon Cloud's CHI@TACC cluster~\cite{keahey2020lessons}.
\autoref{fig:1000-genomes-results} shows workflow stage start and end times for a baseline implementation, which uses Globus Compute's native futures for data synchronization between tasks, and a ProxyFutures implementation.
As each stage can contain up to thousands of tasks, we consolidate the tasks within stages for clarity.
ProxyFutures reduce workflow makespan by 36\%, by better overlapping task execution and communication costs across stages.
More specifically:
(1) tasks within stages 1, 2, and 3 are better overlapped, reducing the stage makespans by 47--48\%;
(2) response time, the time between receiving a task result and submitting another task, is improved (for example, by 54\% when starting stage 4); and
(3) stages 4 and 5 are 5\% faster due to reduced data transfer overheads.
We also note there are no dependencies between tasks within stages 4 or 5, so these stages do not benefit to the same degree as the earlier stages.

\input{figures/deepdrivemd-results}

\label{sec:applications:deepdrivemd}
\textbf{DeepDriveMD:}
We modify the Parsl implementation of DeepDriveMD~\cite{brace2024deepdrivemd} to stream inference batches and results to and from a single, persistent inference task with ProxyStream.
A persistent inference task eliminates task overheads and enables reuse of models and caches.
Streaming with proxies reduces overheads in the DeepDriveMD client because received inference results are immediately added to a queue of simulation task inputs.
In addition to ProxyStream, ProxyFutures are used to indicate availability of a new ML model to the inference task and proxy references for management of intermediate task data.

We compare the performance of DeepDriveMD to a version that uses proxy patterns.
We run each version for three hours using 40 GPUs on Polaris, dedicating one GPU for inference, one for training, and the remainder for simulations.
Round-trip inference time, shown in \autoref{fig:deepdrivemd-results}, is reduced from an average of 21.9$\pm$8.8~s to 15.0$\pm$8.4~s, a 32\% improvement, and 21\% more inference batches were processed in the same wall time.
Reducing inference time is key to enabling greater simulation throughput, such as when the number of simulation workers is increased or simulation time is reduced.

\input{figures/mof-generation-results}

\textbf{MOF Generation:}
We modify the MOF Generation application to communicate all task input and output data larger than 1~kB via proxies.
(The overhead of proxying simple data types such as boolean flags or configuration strings is greater than sending those objects directly.)
We deploy the application with default settings on ten Polaris nodes.
We run the application twice: with the standard proxy implementation of \proxystore{} and with our proxy ownership model.
Here, ownership was sufficient; we did not use the lifetimes model.
We record the number of actively proxied objects during the application's runtime.
As shown in \autoref{fig:mof-generation-results}, the ownership model appropriately evicts proxied data when the lifetime of the associated proxy ends without altering the runtime behaviour of the application.
Manual memory management is possible, as discussed in~\autoref{sec:evaluation:memory}, but automated management is safer and makes adoption of advanced programming practices, such as those we present here, easier and more appealing.

\section{Related Work}
\label{sec:related}

\textbf{Futures} are a pervasive programming abstraction for asynchronous and concurrent programming~\cite{baker1977futures,friedman1978parallel}.
Implicit futures act as references; any dereference blocks automatically until the value is resolved~\cite{reyes2019futures}; thus, they typically require language-level support~\cite{clark1981relational,clark1986parlog,foster1989strand,foster1992productive,chandy1992compositional}.
Explicit futures provide a public interface, such as a \texttt{get} method, that must be invoked to block and retrieve the value;
consequently, they can be provided by languages and third-party libraries.

Explicit futures require control flow synchronization code, which reduces code flexibility and complicates functions that want to operate on a future or a value directly.
Either two implementations or multiple execution paths must be present to support each case.
Implicit futures are also inflexible because they require that the language's type system handle the mechanics of lifting the value out of the future transparently.
Thus few languages support implicit futures, and programmers have limited ability to modify the resolution and lifting processes.
ProxyFutures address these key limitations by providing both an explicit mechanism, the \texttt{Future}, and an implicit mechanism, the \texttt{Proxy}, for Python applications.

Distributed futures represent values that, when available, may be located in remote process memory.
Distributed futures are often underpinned by a remote procedure call (RPC) system, such as in Dask~\cite{rocklin2015dask}, PyTorch~\cite{pytorchrpc}, and Ray~\cite{moritz2018ray,wang2021ownership}.
Because these futures are implemented by the RPC framework, rather than the language, all are necessarily explicit futures, and their use is limited to the confines of the framework.
Thus, for example, one cannot create a distributed future in Dask or Ray and then invoke a serverless function with Globus Compute~\cite{chard20funcx} on that future.
In contrast, ProxyFutures works across frameworks and supports many mediated communication methods via a robust and extensible plugin system.

\textbf{Streaming} applications in which producers and consumers generate and process data continuously are commonly executed at scale on high-performance and cloud computing systems.
Their persistence and resilience needs may be met by message queuing systems such as Apache Kafka~\cite{kafka}, Redis~\cite{redis}, and RabbitMQ~\cite{rabbitmq}.
However, these systems typically optimize for high-throughput, low-latency transmission of small, structured events, in order that these events can be aggregated, filtered, or transformed, as in Kafka.

In contrast, high-performance science applications often produce large raw or unstructured data accompanied by structured metadata~\cite{bicer2017synchrotron}.
File-oriented distributed applications often use GridFTP~\cite{chard2014globus,bryce2012saasglobus}.
Dispel4py~\cite{filguiera2014dispel4py,liang2022dispel4py} maps abstract definitions of streaming workflows onto concrete distributed execution frameworks, such as Python multiprocessing or MPI~\cite{mpi}.
Streamflow~\cite{herath2010streamflow} extends the DAG-based workflow model to integrate continuous event processing.
ADIOS WASP~\cite{choi2016stream}, a data staging platform for scientific stream processing, uses a self-describing file format and supports advanced networking technologies such as RDMA.
The SciStream middleware~\cite{chung2022scistream} enables fast, secure memory-to-memory streaming between nodes that lack direct network connectivity.
CAPIO~\cite{martinelli2023capio} provides a middleware layer for injecting I/O streaming capabilities into file-based workflows.

Consuming an entire stream item (data and metadata) is expensive when only metadata are needed for decision making or data is to be forwarded to another application component.
ProxyStream decouples event metadata notification from bulk data transfer.
Streaming proxies allows data transfers to occur when and where needed, with specifics of the message broker and data storage abstracted from the program.

\textbf{Garbage collection} in distributed environments is challenging.
Automatic techniques such as reference counting and tracing garbage collectors exist, but often requires a priori knowledge by the application programmer to add custom logic for shared object management, and can be inefficient in distributed environments \cite{lerman1986reference,bevan1987distributed,piquer1991indirect,moreau2001construction}.
Maintaining global reference counts or traces adds network overheads, single sources of failure (if reference counting is centralized), or atomicity/consistency challenges (if reference counting is distributed).

Leases, a decentralized, time-based mechanism, can be used to avoid maintaining a shared state across processes~\cite{gray1989leases}.
Task-based execution engines can avoid shared state problems and the complexities of reference count message passing because the central client or scheduler can act as a single source of truth~\cite{rocklin2015dask,moritz2018ray}.
The notion of ownership uses a program's inherent structure to decentralize state management.
In PyTorch RPC, each object has a single owner that maintains the global reference count as remote processes need to access the data~\cite{pytorchrpc}.
Related work extends this concept to implement distributed futures and task recovery in Ray~\cite{wang2021ownership}.

Our proxy-based approach avoids the complexities of global reference counting by associating object lifetimes with tasks, and our framework-agnostic approach means that object scopes can be appropriately managed across complex, distributed applications.

\section{Conclusion}
\label{sec:conclusion}

The lazy object proxy is a powerful construct for building distributed applications, providing benefits of both pass-by-reference and pass-by-value while abstracting low-level communication details from consumers.
Here, we have applied this construct to realize three powerful parallel programming patterns:
a compute framework agnostic distributed futures system, a composable streaming interface for data-intensive workloads, and an ownership model for object management in distributed, task-based applications.
We evaluated these patterns through synthetic benchmarks and showcased three classes of scientific applications that can benefit from the proxy paradigm powered patterns.
Specifically, we reduced the 1000 Genomes workflow makespan by 36\%, reduced DeepDriveMD inference latency by 32\%, and optimized memory usage during MOF generation.

These patterns enable the development of robust, scalable, and portable applications.
For example, ProxyFutures empowers data flow dependencies between tasks executed across different execution engines, such as when one engine is used for local execution on a cluster and another for remote execution on cloud resources.
ProxyStream can support long-running scientific campaigns by using cloud-hosted message brokers for reliable metadata streaming and Globus Transfer for federated, persistent bulk storage and efficient transfer.
The proxy ownership model provides automated wide-area memory management for distributed and cross-site workflows.
In the future, we will investigate further programming patterns that can be enhanced with the proxy paradigm.
Our work here serves as a reference for integrating these design patterns into execution frameworks, such as Dask, Globus Compute, or Parsl, and other high-performance computing toolkits.
By providing first-class support for these patterns directly within commonly used frameworks, we expect to enable speedups in many scientific applications.

\section*{Acknowledgments}
This research was supported in part by the National Science Foundation under Grant 2004894 and the ExaWorks Project and ExaLearn Co-design Center of the Exascale Computing Project (17-SC-20-SC), a collaborative effort of the U.S. Department of Energy Office of Science and the National Nuclear Security Administration.
We used resources provided by the Argonne Leadership Computing Facility (ALCF), a DOE Office of Science User Facility supported under Contract DE-AC02-06CH11357, and the Chameleon testbed supported by the National Science Foundation.

\balance
\bibliographystyle{IEEEtran}
\bibliography{refs}

\end{document}

%% file: figures/patterns-overview.tex
\begin{figure}
    \centering
    \includegraphics[width=\columnwidth,trim={0 0px 0 0px},clip]{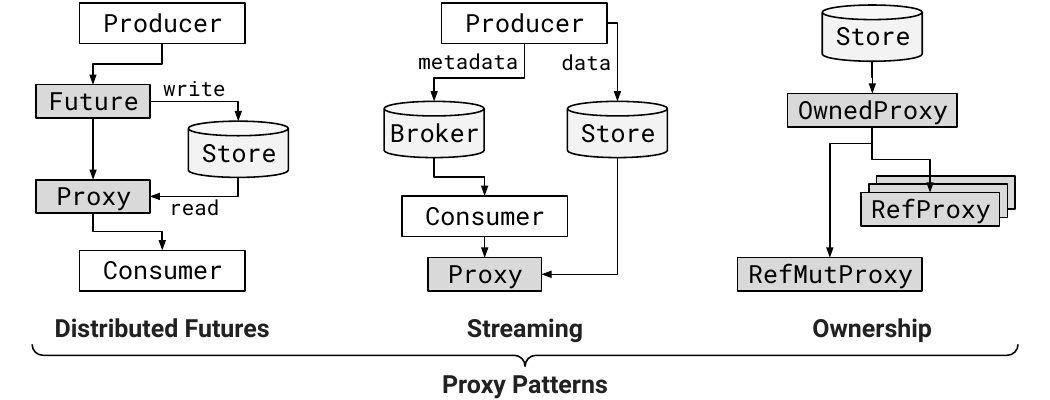}
    \caption{
        Overview of the three proxy-based data flow patterns we design.
    }
    \label{fig:patterns-overview}
\end{figure}

%% file: figures/patterns-proxystore-stack.tex
\begin{figure}[t]
    \centering
    \includegraphics[width=\columnwidth,trim={2mm 0px 2mm 0px},clip]{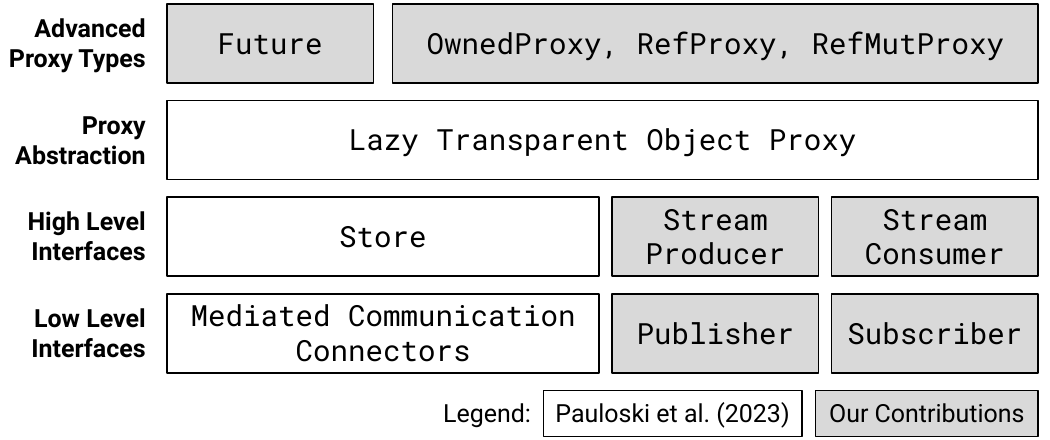}
    \caption{
        Overview of the ProxyStore interface and abstraction stack with our contributions included in the shaded boxes.
    }
    \label{fig:patterns-proxystore-stack}
\end{figure}

%% file: figures/task-pipelining.tex
\begin{figure}[t]
    \centering
    \includegraphics[width=\columnwidth,trim={0 0px 0 0px},clip]{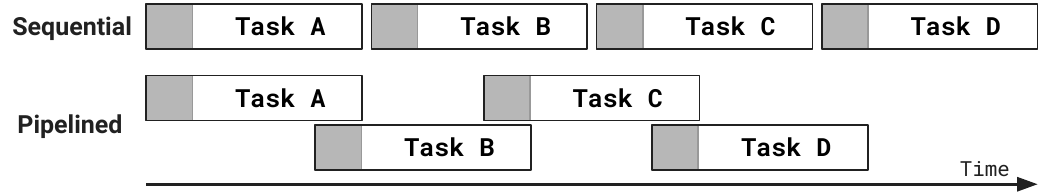}
    \caption{
        Four tasks executed in a sequential (above) or pipelined (below) fashion. Each task produces data needed by the following task. The grey region at the start of each task represents startup overhead before the input data can be used. By enabling a successor task to start before its predecessor has finished, futures enable overlapping of startup overhead with computation, a form of pipelining.
    }
    \label{fig:task-pipelining}
\end{figure}

%% file: listings/futures-example.tex
\lstset{escapeinside={(*}{*)}}
\begin{lstlisting}[style=PythonStyle, label={lst:futures-example}, caption={Example usage of the ProxyFuture interface within tasks executed by Dask. A proxy created from a \texttt{Future} will block implicitly on the result of the future when needed. This interface abstracts the low-level communication away from the functions which set the result or consume the proxy.}, float, floatplacement=t]
from dask.distributed import Client
from proxystore.connectors.foo import FooConnector
from proxystore.store import Store
from proxystore.store.future import Future

def producer(future: Future[str]) -> None:
    future.set_result('value')

def consumer(data: Proxy[str]) -> None:
    assert data is 'value'

with Store('example', FooConnector()) as store:
    client = Client(...)
    future: Future[str] = store.future()

    t1 = client.submit(producer, future)
    t2 = client.submit(consumer, future.proxy())

    t1.result(), t2.result()
\end{lstlisting}

%% file: figures/proxystore-streaming.tex
\begin{figure}
    \centering
    \includegraphics[width=\columnwidth,trim={0 0px 0 0px},clip]{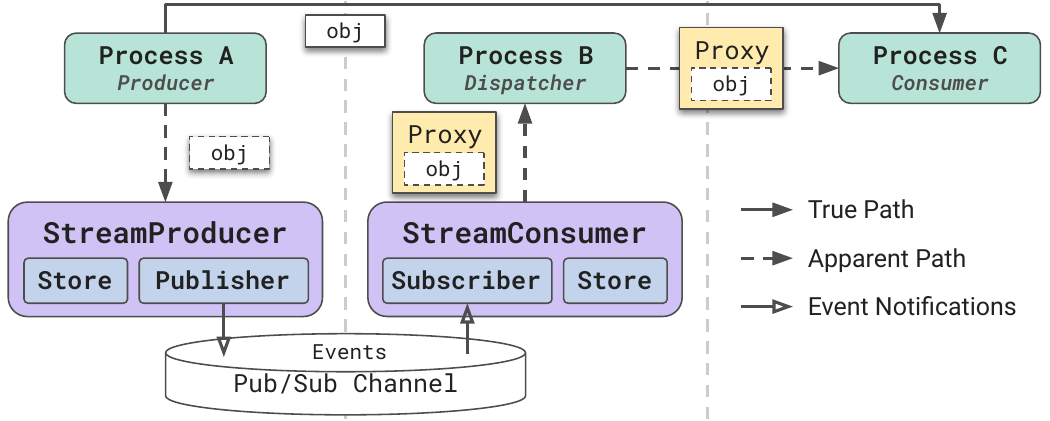}
    \caption{
        The \texttt{StreamProducer} abstracts low-level communication details from the \texttt{StreamConsumer} and transparently decouples metadata from bulk data transfer. Yielding proxies, rather than objects directly, in the \texttt{StreamConsumer} enables just-in-time resolution and pass-by-reference optimizations.
    }
    \label{fig:proxystore-streaming}
\end{figure}

%% file: listings/streaming-globus-compute.tex
\lstset{escapeinside={(*}{*)}}
\begin{lstlisting}[
    style=PythonStyle,
    label={lst:streaming-gc},
    caption={
    Example using the ProxyStream interfaces to stream data between two tasks executed remotely using Globus Compute.
    A Kafka broker is used for metadata and an arbitrary \texttt{FooConnector} for bulk data transfer.},
    float,
    floatplacement=t,
]
from globus_compute_sdk import Executor
from proxystore.stream import (StreamProducer,
    StreamConsumer, KafkaPublisher, KafkaSubscriber)

def producer() -> None:
    store = {'topic': Store(..., FooConnector(...))}
    publisher = KafkaPublisher(...)
    
    with StreamProducer(publisher, store) as producer:
        for item in ...:
            producer.send('topic', item) 

def consumer() -> None:
    subscriber = KafkaSubscriber('topic', ...)  
    
    with StreamConsumer(subscriber) as consumer:
        for item in consumer: 
            assert isinstance(item, Proxy)

with Executor('<Endpoint UUID>') as client:
    t1 = client.submit(producer)
    t2 = client.submit(consumer)

    t1.result(), t2.result()
\end{lstlisting}

%% file: listings/ownership-stubs.tex
\lstset{escapeinside={(*}{*)}}
\begin{lstlisting}[
    style=PythonStyle,
    label={lst:ownership-stubs},
    caption={
    Proxy ownership model interfaces and functions.
    Functions are preferred over methods on the associated proxy reference types to prevent unintentionally clobbering a method of the same name on the target object.
    },
    float,
    floatplacement=t,
]
class Store(Generic[Connector]):
    def owned_proxy(obj, ...) -> OwnedProxy: ...

def into_owned(Proxy) -> OwnedProxy: ...
def borrow(OwnedProxy) -> RefProxy: ...
def mut_borrow(OwnedProxy) -> RefMutProxy: ...
def clone(OwnedProxy) -> OwnedProxy: ...
def update(OwnedProxy | RefMutProxy) -> None: ...
\end{lstlisting}

%% file: listings/lifetimes.tex
\lstset{escapeinside={(*}{*)}}
\begin{lstlisting}[
    style=PythonStyle,
    label={lst:lifetimes},
    caption={Example usage of lifetimes when creating a proxy. A \texttt{Lifetime} instance represents a physical or logical scope that will clean up all resources (i.e., objects) that were associated with the lifetime when closed.},
    float,
    floatplacement=t
]
from proxystore.connectors import FooConnector
from proxystore.store import Store
from proxystore.store.lifetimes import LeaseLifetime

with Store('example', FooConnector()) as store:
    lease = LeaseLifetime(store, expiry=10)  
    proxy = store.proxy('value', lifetime=lease)
    lease.extend(5)  
    time.sleep(20)  
    assert lease.done()  
    # Object associated with the proxy has been removed
\end{lstlisting}

%% file: figures/task-pipelining-results.tex
\begin{figure}
    \centering

    \subfloat[Task schedules for no-proxy; proxy; ProxyFutures $f=0.2$, $f=0.5$.]{
        \includegraphics[width=\columnwidth,trim={0 2px 0 2px},clip]{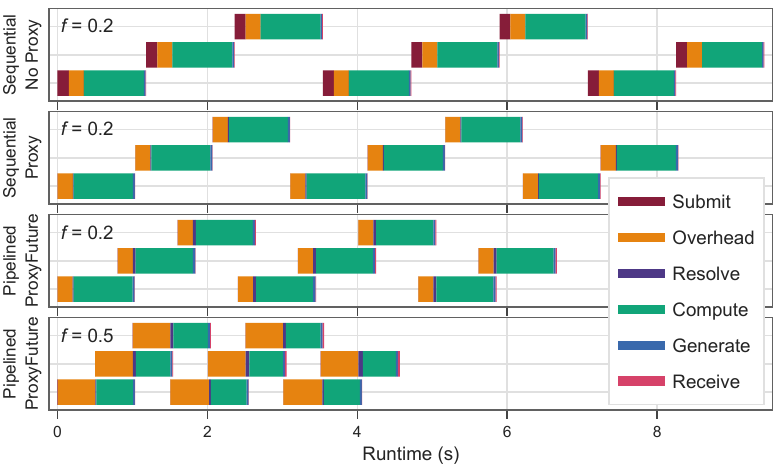}
        \label{fig:task-pipelining-timeline}
    }

    \subfloat[Makespan vs.\ overhead fraction for no-proxy, proxy, ProxyFutures.]{
        \includegraphics[width=\columnwidth,trim={0 2px 0 0px},clip]{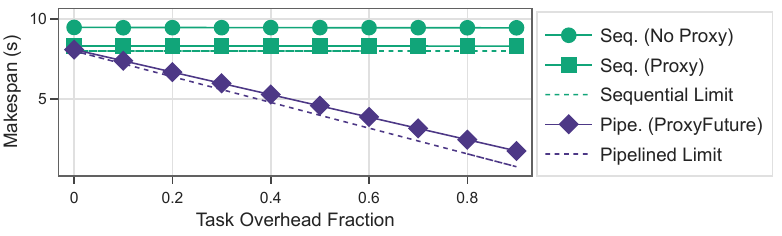}
        \label{fig:task-pipelining-trend}
    }

    \caption{
        Results for synthetic benchmark with 8 tasks, each sleeping for 1~s and communicating 10~MB to its successor, and with overhead fraction $f$ determining how much of the 1~s can be overlapped with its predecessor task.
        (Top) Task execution schedules in four scenarios: \textit{sequential no proxy}, with delays due to workflow engine submission costs; 
        \textit{sequential proxy}, with proxies enabling immediate task start after proxy is resolved; and two \textit{pipelined ProxyFuture} cases ($f=0.2$ and $f=0.5$), in which distributed futures relax strict inter-task dependencies and enable pipelining to overlap initial task overheads.
        The \emph{overhead} and \emph{compute} sleeps dominate in all cases, while
        times to \emph{resolve} task input data and \emph{receive} task results increase, with overhead fraction, while makespan decreases due to pipelining overlap.
        (Bottom) Synthetic benchmark makespan vs.\  overhead fraction, for no proxy, proxy, and ProxyFuture scenarios.
        Each value is averaged over five runs;
        standard deviations are all less than 20~ms.
    }
    \label{fig:task-pipelining-results}
\end{figure}

%% file: figures/streaming-results.tex
\begin{figure*}
    \centering
    \includegraphics[width=\textwidth,trim={0 0px 0 0px},clip]{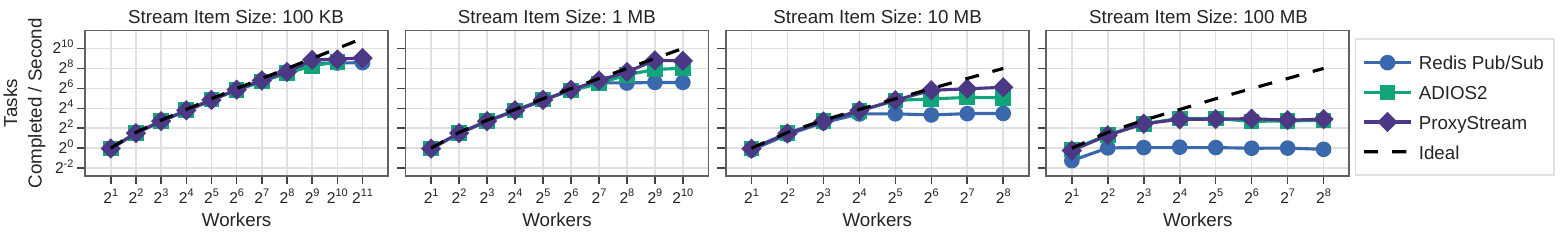}
    \caption{
        Compute tasks completed per second as a function of stream item data size and number of workers.
        One worker generates data consumed by a central dispatcher that launches simulated compute tasks (one second sleep tasks) for each item across the remaining $n-1$ workers.
        At small data sizes ($\leq 100$~KB), data transfer overheads are negligible and the dispatcher can keep up with incoming stream data; however, at large data sizes and worker counts, the dispatcher becomes overwhelmed by the size of data transfers required for each task in the Redis Pub/Sub configuration.
        ProxyStream transparently decouples data flow from control flow improving overall system performance as stream data sizes and the number of workers is increased.
        }
    \label{fig:streaming-results}
\end{figure*}

%% file: figures/ownership-results.tex
\begin{figure}
    \centering
    \includegraphics[width=\columnwidth,trim={0 0px 0 0px},clip]{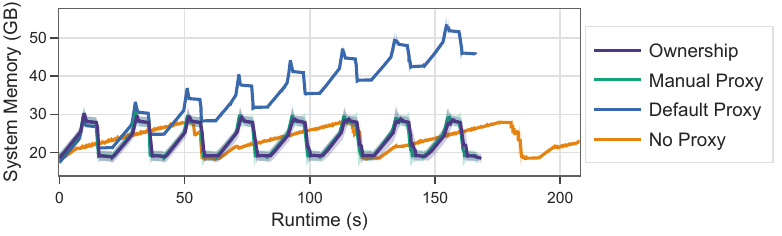}
    \caption{  
        Average system memory usage over three runs of a simulated MapReduce workflow.
        Shaded regions denote standard deviation in memory usage.
        Memory management limitations in ProxyStore cause baseline memory utilization to increase over time.
        Manual management can alleviate this problem, but requires careful implementation and prior knowledge.
        In contrast, our ownership model provides automated memory management equal to a hand-tuned implementation and enforces a set of rules at runtime.
    }
    \label{fig:ownership-results}
\end{figure}

%% file: figures/1000-genomes-results.tex
\begin{figure}
    \centering
    \includegraphics[width=\columnwidth,trim={0 0px 0 0px},clip]{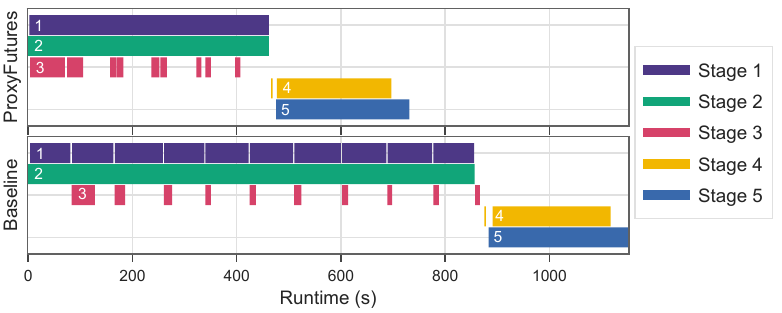}
    \caption{
        1000 Genomes Workflow stage start and ends times.
        ProxyFutures reduces workflow makespan by starting computations when data are available rather than when prior tasks complete. 
    }
    \label{fig:1000-genomes-results}
\end{figure}

%% file: figures/deepdrivemd-results.tex
\begin{figure}
    \centering
    \includegraphics[width=\columnwidth,trim={0 0px 0 0px},clip]{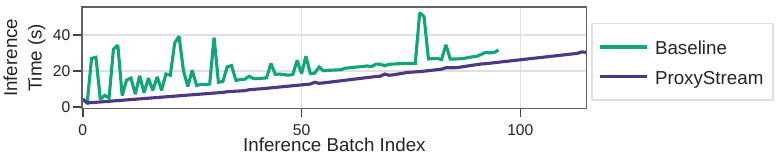}
    \caption{
        Comparison of inference round-trip time between two DeepDriveMD implementations: baseline and ProxyStream. The size of each batch increases over time as the application accumulates more data points.
    }
    \label{fig:deepdrivemd-results}
\end{figure}

%% file: figures/mof-generation-results.tex
\begin{figure}
    \centering
    \includegraphics[width=\columnwidth,trim={0 0px 0 0px},clip]{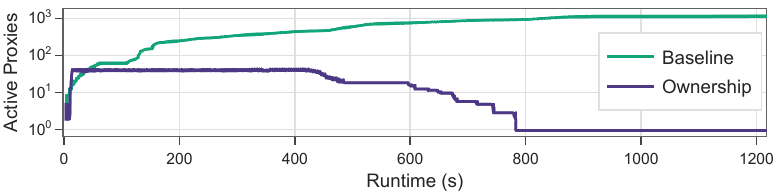}
    \caption{
        Number of active proxies (i.e., proxies that still have a stored target object) during the runtime of the MOF Generation application.
        Our ownership model for proxies appropriately cleans up proxies when no longer needed while maintaining the benefits of the pass-by-reference model.
    }
    \label{fig:mof-generation-results}
\end{figure}